\documentclass[a4paper,showpacs,nofootinbib]{revtex4}
\usepackage[utf8]{inputenc}
\usepackage{graphicx,bm,color,amssymb,amsmath}
\usepackage{slashed,verbatim}
\usepackage{subfigure}
\usepackage[colorlinks=true,linktocpage=true,linkcolor=blue,citecolor=blue]{hyperref}

\begin{document}
\title{Deciding on the anomalous magnetic moment of quarks in a framework of nonlocal NJL model}
\author{Chowdhury Aminul Islam$^{1,a}$, Mahammad Sabir Ali$^{2,3,b}$, Mei Huang$^{1,c}$}
\affiliation{$^1$ School of Nuclear Science and Technology, University of Chinese Academy of Sciences, Beijing 100049, China}
\affiliation{$^2$ Department of Theoretical Physics, Tata Institute of Fundamental Research, Homi Bhabha Road, Mumbai 400005, India}
\affiliation{$^3$ School of Physical Sciences, National Institute of Science Education and Research, HBNI, Jatni, Khurda 752050, India}
\email{$^a$ chowdhury.aminulislam@ucas.ac.cn}
\email{$^b$ sabir@niser.ac.in}
\email{$^c$ huangmei@ucas.ac.cn}

\begin{abstract}
{Anomalous magnetic moment (AMM) of quarks in presence of an external magnetic field has been explored using a nonlocal Nambu\textemdash Jona-Lasinio (NJL) model. Various strengths of AMM differing in orders of magnitude are used in the literature. We explore them in our nonlocal framework to decide on their strength. We checked the validity of using constant AMM of quarks and investigate two different temperature and magnetic field dependent forms. The forms are taken as the AMM being proportional to the i) meanfield and the ii) square of the meanfield.  On comparison with the lattice data for both the condensate averages and differences, it turns out that the second choice is the most suitable one, in such an effective model scenario. In the process, we also keep track of the phase diagram in the $T-eB$ plane arising from our model calculation. The outcome is reasonable as far as the lattice QCD result is concerned.}
\end{abstract}

\maketitle

%%%%%%%%%%%%%%%%%%%%%%%%%%%%%%%%%%%%%%%%%%%%%%%%%%%%%%%%%%%%%%%%%%%%%%%%%%%                         
\section{Introduction}
\label{sec:intro}
Quantum chromodynamics (QCD) matter under an external magnetic field has attracted wide interest for the scenarios of the early universe, neutron stars and of course the heavy ion collisions. It shows a lot of novel phenomena, e.g. the chiral magnetic effect(CME)~\cite{Kharzeev:2007tn,Kharzeev:2007jp,Fukushima:2008xe,Kharzeev:2010gr}, the magnetic catalysis (MC) in the vacuum ~\cite{Klevansky:1989vi,Klimenko:1990rh,Gusynin:1995nb}, the inverse magnetic catalysis (IMC) around the critical temperature of the chiral phase transition~\cite{Bali:2011qj,Bali:2012zg,Bali:2013esa} etc. Besides, lattice results also show that the magnetized QCD matter behaves as diamagnetic (negative susceptibility) at low temperatures and paramagnetic (positive susceptibility) at high temperature~\cite{Bali:2012jv,Bali:2020bcn}. For reviews see  Refs.~\cite{Andersen:2014xxa,Miransky:2015ava,Huang:2015oca,Kharzeev:2015znc,Bzdak:2019pkr}.

Hadronic properties also change in the presence of a magnetic field, as the latest lattice calculations show that  the neutral pion mass decreases with the magnetic field and then saturates~\cite{Bali:2017ian}. In contrast, the charged pion mass shows a non-monotonic magnetic field-dependent behaviour: first increases and then decreases as a function of the magnetic field ~\cite{Ding:2020jui}. 

There have been some efforts to understand the mechanism(s) behind all those manifestations of QCD matter in presence of a magnetic field. For example, in the local version of the Nambu\textemdash Jona-Lasinio (NJL) model, the neutral pion fluctuation\cite{Fukushima:2012kc}, the chirality imbalance \cite{Chao:2013qpa,Yu:2014sla} and the running coupling with the magnetic field~\cite{Farias:2014eca,Ferreira:2014kpa} have all been investigated in great details to understand the IMC behaviour. On the other hand, the non-local version of the model can lead to the IMC effect naturally, without introducing an explicit magnetic field dependent four Fermi interaction~\cite{Pagura:2016pwr}. 

This leaves us with the choice of working with less number of parameters. The non-local version captures the essence of the running of the QCD coupling constant, in contrast to its local counterpart. In the lattice calculations, it is argued that gluons and the sea quarks through which the gluons feel the presence of the magnetic field plays the most important role in giving rise to the IMC effect~\cite{Bruckmann:2013oba}. We also know that these extra interactions (as compared to the quantum electrodynamics) among the gluons themselves give rise to the asymptotic nature of QCD. In a nonlocal NJL model, one gets the same qualitative behaviour through the nonlocal form factor which gives rise to IMC effect around the transition temperature.

It is known from the basic perturbation theory that the free massless quarks cannot have measurable anomalous magnetic moments (AMM). With the dynamical breaking of the chiral symmetry rendering the quarks massive, measurable strengths of AMM of quarks are generated~\cite{Bicudo:1998qb,Chang:2010hb}. Thus, the AMMs of quarks are generated dynamically.

Chiral symmetry breaking in QCD, being a nonperturbative phenomenon, cannot be dealt with first principle analytical methodology. To understand such nonperturbative aspects of QCD, particularly related to the global symmetries, effective models such as NJL play a quite handy role~\cite{Klevansky:1992qe,Hatsuda:1994pi}. In the same spirit, AMM of quark has also been investigated using such effective model~\cite{Bicudo:1998qb}. Recently, consideration of the AMMs of quarks in effective model scenarios has reignited, particularly with the discovery of the IMC effect by the lattice QCD~\cite{Bali:2011qj,Bali:2012zg}. There have been a series of works exploring the effects~\cite{Fayazbakhsh:2014mca,Chaudhuri:2019lbw,Chaudhuri:2020lga,Xu:2020yag,Mei:2020jzn,Aguirre:2020tiy,Chaudhuri:2021skc,Ghosh:2021dlo,Aguirre:2021ljk,Wen:2021mgm,Farias:2021fci,Wang:2021dcy,Chao:2022bbv,Mao:2022dqn,He:2022inw,Qiu:2023kwv}. 

In Ref.~\cite{Fayazbakhsh:2014mca}, the authors considered three different sets of AMM of quarks and determined the constituent quark masses. They observed the occurrence of IMC for large values of AMM. Considering the dynamically generated AMM to be proportional to the meanfields, Ref.~\cite{Xu:2020yag} found that the constituent quark mass decreases with increasing values of AMM. They have further explored the magnetic susceptibility and meson spectrum in presence of AMM. In Ref.~\cite{Ghosh:2021dlo}, the authors have tried to understand the AMM of quark by considering the thermomagnetic modification through the effective photon-quark-antiquark vertex function in an NJL model. In a recent study~\cite{Kawaguchi:2022dbq}, a detailed analysis of the phase transition in presence of a magnetic field, within a local NJL model, has been carried out considering different functional forms of AMM.

In this article, we study the effect of anomalous magnetic moment (AMM) of quarks within the setup of non-local NJL. So far, to the best of our knowledge, all the effective model studies involving quarks' AMM have been done in the local version of the model. Most of these works introduce AMM to obtain IMC, whereas, as mentioned above, IMC arises naturally in the non-local NJL model. Thus it provides a good opportunity to explore the AMM of quarks in a non-local setup.

Though AMM of quarks has been extensively studied there is no consensus on their strengths. Even there could be two orders of magnitude differences between two sets of values depending on the choices of the parameters~\cite{Fayazbakhsh:2014mca}. Then there is also the issue of using the constant values of AMM of quarks. We know from the perturbative calculation that a massless fermion cannot possess a measurable AMM~\cite{Bicudo:1998qb,Chang:2010hb}. The dynamical breaking of chiral symmetry generates AMM for quarks. We also know that at high enough temperature ($\approx 150$ MeV) the chiral symmetry gets restored and therefore the generated measurable AMM of quarks should also vanish. This understanding clearly indicates that using a constant value for the AMM of quark may bear unjustified implications.

Thus, we think that there are mainly two issues pertaining to the use of AMM of quarks in effective models. One is the strength of the AMM and the other is its dependence on external parameters like temperature. In fact, there are some studies which have argued and used a temperature and magnetic field dependent AMM~\cite{Xu:2020yag,He:2022inw,Kawaguchi:2022dbq}.

In our study, we explore both the constant and the temperature dependent AMM. The temperature dependence comes through the ansatz that the strength of the AMM depends on the strength of the meanfield. We take two different ansatzes \textemdash\, it is proportional to the i) meanfield and ii) square of the meanfield with different strengths of the proportionality constant. The second choice is particularly inspired by the Ref.~\cite{Lin:2021bqv}.
%\cite{Kawaguchi:2022dbq}
In deciding the strength of the AMM or choosing the appropriate functional form of its temperature dependence our goal is simple. We rely on the first principle lattice QCD calculation for the condensate average and condensate difference in presence of the magnetic field~\cite{Bali:2012zg}. The same quantities are then calculated within the ambit of the model and compared with the available LQCD data. The phase diagram in the $T-eB$ plane plays an important role as well.

The behaviour of the condensate average as a function of temperature for different values of magnetic field decides how well the IMC effect is captured. This helps us to decide on the strength as well as the form of the AMM. This remains our first criterion. We explore the second criterion with the strength and the form which fulfills the first criterion. The data on the condensate difference for different values of the magnetic field from LQCD calculation is our second criterion. We compare the model predictions for the condensate difference with these lattice QCD data.

Thus we look for the strength and the form of the AMM which produce comparable values for both the averages and differences of condensate at different values of the magnetic field, as given by the lattice QCD. Using the values of the condensates the phase diagram in $T-eB$ plane we finally decide on the form and the strength of the AMM of quarks. Our conclusion is that the strength and the form with a better agreement with LQCD data are better suited to be used within such model premises.

The manuscript is organised as follows. In Sec.~\ref{sec:formalism}, we give the detail of the model that we use. Particularly, we introduce the AMM in the non-local NJL model. Then, in Sec.~\ref{sec:res}, we present our result and finally we conclude in Sec.~\ref{sec:con}.

\section{Formalism}
\label{sec:formalism}
Here, we briefly discuss our formalism. We have used nonlocal NJL model for our calculation and the details can be found in Ref.~\cite{Ali:2020jsy}. The 2-flavour nonlocal NJL Lagrangian in presence of a uniform magnetic field is given as
\begin{align}
{\cal L} = \bar{\psi}(x)\left(i\gamma^\mu D_\mu-m_0\right)\psi(x)+{\cal L}_1+{\cal L}_2,
\end{align}
where, the covariant derivative is given by $D_\mu=\partial_\mu+ie\hat{Q}A_\mu^{ext}$ with $\hat{Q}$ being the diagonal charge matrix with $u$ and $d$ quark charges as the diagonal elements. Though the isospin symmetry is explicitly broken by the magnetic field due to unequal electric charges of $u$ and $d$ quarks, we will consider $m_0=m_c\mathbb{I}_{2\times2}$. $A_\mu^{ext}$ is the electromagnetic gauge field, which in our case is the external magnetic field. The interaction terms ${\cal L}_1$ and ${\cal L}_2$ are given as~\cite{Bowler:1994ir,Plant:1997jr,General:2000zx,Praszalowicz:2001wy,GomezDumm:2001fz,GomezDumm:2006vz,Pagura:2016pwr,GomezDumm:2017iex},
\begin{eqnarray}
{\cal L}_1&=&G_1\left\{j_a(x)j_a(x)+\tilde{j}_a(x)\tilde{j}_a(x)\right\}\,\, {\mathrm {and}}\nonumber\\
{\cal L}_2&=&G_2\left\{j_a(x)j_a(x)-\tilde{j}_a(x)\tilde{j}_a(x)\right\},
\label{eq:nl_l}
\end{eqnarray}
respectively.  The nonlocal currents $j_{a}(x)$ and $\tilde{j}_a(x)$ are given by~\cite{GomezDumm:2006vz,Ali:2020jsy}
\begin{eqnarray}
j_{a}(x)/\tilde{j}_a(x)=\int d^4z\ {\cal H}(z)\bar{\psi}\left(x+\frac{z}{2}\right)\Gamma_{a}/\tilde{\Gamma}_a\psi(x-\frac{z}{2}).
\label{eq:current}
\end{eqnarray}
Here, $\Gamma_a=(\mathbb{I},i\gamma_5\vec{\tau})$,
$\tilde{\Gamma}_a=(i\gamma_5,\vec{\tau})$ with $\tau$'s being the Pauli matrices, $a=\{0,1,2,3\}$ and ${\cal H}(z)$ is the non-local form
factor in the position space. It is to be mentioned here that the symmetry of the term ${\cal L}_1$ is $SU(2)_V\times SU(2)_A\times U(1)_V\times U(1)_A$, it is the ${\cal L}_2$ term which breaks the $U(1)_A$ symmetry explicitly.

To integrate out the fermionic degrees of freedom, one needs to linearise the Lagrangian, which is also known as bosonisation. During linearisation one can introduce four auxiliary fields associated with four different types of interactions. The details of such bosonisation procedure can be found in the appendix of Ref.~\cite{Hell:2008cc}. Within meanfield approximation, these auxiliary fields will get nonzero expectation values allowed by symmetries.

If there is no isospin breaking external agents like isospin chemical potential $(\mu_I)$ or magnetic field, then with $m_u=m_d$ we have the nonzero meanfield associated with the current $j_1(x)$, which breaks the $SU(2)_A$ symmetry. And one can easily notice that it depends only on the combination $G_1+G_2$. But, if we have any isospin breaking effects, then $SU(2)_V$ is also explicitly broken and one can have nonzero expectation value associated with the current $j_{\Vec{a}(x)}$. This new meanfield depends on the combination of $G_1-G_2$. On the other hand, there will be no pseudoscalar condensates because of the parity conservation. To keep things general, the coupling constants $G_1$ and $G_2$ can be parameterized in the following manner
\begin{eqnarray}
\nonumber
G_1&=&(1-c)G_0/2\\
G_2&=&cG_0/2\;,
~\label{eq:cdef}
\end{eqnarray}
with $c=1/2$ corresponding to the standard NJL model.

We denote the isoscalar and isovector fields as $\sigma$ and $\pi$, respectively, with `$s$' and `$ps$' in the subscripts representing the Lorentz scalar and pseudoscalar, respectively. Since we will be working in presence of $eB$, the isospin is explicitly broken and there will be nonzero values for the condensate, $\langle\bar{\psi}(x)\vec{\tau}\psi(x)\rangle$.

For the ease of the readers, we briefly discuss here the formalism of the case in which $\sigma_s$ is the only condensate that survives. The mean field Lagrangian, in that case, reads as,
\begin{align}
{\cal L}_{\rm MF}=&\bar{\psi}(x)\left\{\delta^4(x-y) (-i\slashed{\partial}+m)+{\cal H}(x-y)\sigma_s\left(\frac{x+y}{2}\right)\right\}\nonumber\\
&\times\psi(y)+\frac{1}{2G_0}\sigma_s^2\left(\frac{x+y}{2}\right),
\label{eq:l_mf}
\end{align} 
with the auxiliary field, $\sigma_s$ being given by,
\begin{equation}
\sigma_s(x)=-\, \frac{G_0}{2}\int d^4z\ {\cal H}(z)\bar{\psi}\left(x+\frac{z}{2}\right)\psi(x-\frac{z}{2}).
~\label{eq:sigmasdef}
\end{equation}
From the mean field Lagrangian in Eq.~\ref{eq:l_mf}, the free energy per unit volume can be calculated as,
\begin{align}
\Omega=\frac{S_{\rm MF}}{V^{(4)}}=
-2N_fN_c\int \frac{d^4q}{(2\pi)^4}\ln\left[q^2+M^2(q)\right]
+\frac{\sigma_s^2}{2G_0},~\label{eq:SMFB0}
\end{align}
with the constituent quark mass reads as,
\begin{equation}
M(q)=m+h(q,q)\sigma_s\;.
~\label{eq:MconstB0}
\end{equation}
Here, $h(p,p')$ is the non-local form factor in the momentum space. It is the Fourier transform of ${\cal H}(x-y)$. We choose the form factor to be Gaussian~\cite{Ali:2020jsy},
\begin{equation}
h(p,p')=e^{-(p+p')^2/(4\Lambda^2)}.
\end{equation}
Using the form factor the self-consistent gap equations can be written as
\begin{eqnarray}
\sigma_s = 8 N_c \ G_0 \int \frac{d^4 q}{(2 \pi)^4}\  h(q,q) \ \frac{M(q)} {q^2 + M^2(q)}\;.
\label{eq:gapeq}
\end{eqnarray}
Having obtained $\sigma_s$, one can calculate the condensate by differentiating the potential in Eq.~\ref{eq:SMFB0} with respect to the current quark mass,
\begin{eqnarray}
\langle\bar{\psi}_f(x)\psi_f(x)\rangle=\frac{\partial\Omega}{\partial m} &=& - \, 4 N_c \int \frac{d^4 q}{(2 \pi)^4}\
\frac{M(q)} {q^2 + M^2(q)} \ \ .
\label{eq:cond0}
\end{eqnarray}
The above integral is finite in the chiral limit, as $M(p)$ in Eq.~\ref{eq:MconstB0} approaches zero at large momentum in that limit. But away from the chiral limit the integral is infinite and needs to be regularised. It is done by subtracting an identical term from the above Eq.~\ref{eq:cond0} with the effective mass, $M$ being replaced by current mass, $m$. The technical details can be found in Ref.~\cite{Pagura:2016pwr}.

\subsection{With nonzero AMM of quarks}
\label{ssec:nonzero_amm}
The Lagrangian of a 2-flavor NJL model in presence of a uniform magnetic field with non-zero anomalous magnetic moment (AMM) of the constituent quarks is given by
\begin{align}
{\cal L} = \bar{\psi}(x)\left(i\gamma^\mu D_\mu-m_0+\frac{1}{2}\hat{a}\sigma^{\mu\nu}F_{\mu\nu}\right)\psi(x)+{\cal L}_1+{\cal L}_2,
\end{align}
where the electromagnetic field tensor is given by $F_{\mu\nu}=\partial_\mu A_\nu^{ext}-\partial_\nu A_\mu^{ext}$ and $\sigma^{\mu\nu}=\frac{i}{2}\left[\gamma^\mu,\gamma^\nu\right]$. $\hat{a}$ is the diagonal matrix in flavour space with elements proportional to the respective AMM of quarks. ${\cal L}_1$ and ${\cal L}_2$ are given by Eq.~\ref{eq:nl_l}. In the presence of a magnetic field, which breaks the isospin symmetry explicitly, we have another auxiliary field, $\pi_s$ along with the usual $\sigma_s$~\cite{Boomsma:2009yk}.

With those two auxiliary fields, we can write down the effective Euclidean action using the
mean-field Lagrangian as,
\begin{align}
S_{\mathrm{bos}}=-\ln\det\mathcal{D}+\frac{1}{2G_0}
\int d^{4}x\ \sigma_s^2(x)+\frac{1}{2(1-2c)G_0}\int d^{4}x\ \vec{\pi}_s(x)\cdot\vec{\pi}_s(x).
\end{align}
The fermionic determinant in the above equation is given by [in Landau gauge, with $A_\mu^{ext}=(0,0,Bx_1,0)$],
\begin{align}
\mathcal{D}\left(  x+\frac{z}{2}\,,x-\frac{z}{2}\right)=\,\gamma_{0}\;W\left(  x^+,x\right)  \gamma_{0}
\, \bigg[\,\delta^{(4)}(z)\,\big(-i{\slashed D}+m+\sigma^{12}\hat{a}B\big)\mathcal{H}(z)\big[\sigma_s(x)+\vec{\tau}\cdot\vec{\pi}_{s}(x) \big]
\bigg]\; W\left(  x,x^-\right),
\label{eq:ferm_det}
\end{align}
with $x_{\pm}=x\pm z/2$ and $W(x,y)$ is given as $W(x,y) = \mathrm{P}\exp\left[ -\, i \hat Q\int_{x}^{y}dr_{\mu}\  \mathcal{A}_{\mu }(r)\right]$. The auxiliary fields are assigned with the space-time independent meanfield values just like the non-magnetic field scenario. To make the exploration simpler, without loss of generality, we can choose $\tau^3$ to be the direction of $\vec\pi_s$ ($\pi^3_s$). With other auxiliary fields being zero the fermionic determinant 

As for the non-magnetic field scenario here also we will assign space-time independent meanfield values to the auxiliary fields. Without loss of generality, we can choose $\vec\pi_s$ to be in the $\tau^3$ ($\pi^3_s$) direction. As already mentioned, all other pseudo-scalar auxiliary fields are chosen to have zero mean field values. Then the fermionic determinant and the action are written as,
\begin{align}
\mathcal{D}^{\mbox{\tiny MFA}} (  x , x')= \delta^{(4)}(x-x') \left( - i {\slashed D}
- \hat Q \, B \, x_1\gamma_2 + m+\sigma^{12}\hat{a}B \right)+\left( \sigma_s +\tau_3{\pi}^3_s\right) \mathcal{H}(x-x') \; \exp\left[ \frac{i}{2} \, \hat Q \, B \, (x_2 - x_2')\, (x_1 +
x_1')\right]
\end{align}
and
\begin{equation}
S_{\mathrm{bos}}=-\ln\det\mathcal{D}^{\mbox{\tiny MFA}} +\frac{1}{2G_0}
\int d^{4}x\ \sigma_s^2+\frac{1}{2(1-2c)G_0}
\int d^{4}x\ \left(\pi^3_s\right)^2,
\end{equation}
respectively. The inverse of the propagator in the momentum space can be obtained using the Ritus eigenfunctions~\cite{Ritus:1978cj} using the following transformation
\begin{equation}
	{\cal D}_{p,p'}=\int d^4x\, d^4x' \bar{\mathbb{E}}_p(x){\cal D}(x,x')\mathbb{E}_{p'}(x').
\end{equation}
And we obtain it as,
\begin{widetext}
\begin{eqnarray}
	{\cal D}_{p,p'}&=&(2\pi)^4\delta_{kk'}\delta(p_2-p'_2)\delta(p_3-p'_3)\delta(p_4-p'_4)\bigg\{\left[\mathbb{I}+\delta_{k0}\left(\Delta^{s_f}-\mathbb{I}\right)\right]\left(-s_f\sqrt{2k|qB|}\gamma_2+p_3\gamma_3+p_4\gamma_4-i\hat{a}B\gamma_1\gamma_2\right)\nonumber\\
	&&+\sum_{\lambda=\pm1}\Delta^\lambda M_{\bar{p},k}^{\lambda,f}\bigg\}.
\end{eqnarray}
After performing the determinant operation, the free energy becomes 
\begin{eqnarray}
	\Omega=\frac{S^{\mbox{\tiny MFA}}_{\mathrm{bos}}}{V^{(4)}} & = & \frac{\sigma_s^2}{2 G_0} +\frac{({\pi}^3_s)^2}{2(1-2c)G_0}- N_c \sum_{f=u,d} \frac{  |q_f B|}{2 \pi} \int \frac{d^2
		q_\parallel}{(2\pi)^2} \ \Bigg\{ \ln\left[q_\parallel^2 + \left({M^{s_{\! f},f}_{q_\parallel,0}}+s_fa_f\right)^2\right]
	+ \nonumber \\
	& & \sum_{k=1}^\infty \ \ln\left[ \left\{2 k |q_f B| + q_\parallel^2 +
	\left(M^{+1,f}_{q_\parallel,k}+a_f\right)\left( M^{-1,f}_{q_\parallel,k}-a_f\right)\right\}^2 \! \! + q_\parallel^2 \left(
	M^{+1,f}_{q_\parallel,k} - M^{-1,f}_{q_\parallel,k}+2a_f\right)^2\right]\Bigg\} .
	\label{eq:s_mfa}
\end{eqnarray}
\end{widetext}
With a Gaussian nonlocal form factor, the constituent mass is given by~\cite{Pagura:2016pwr},
\begin{equation}
M^{\lambda,f}_{q_\parallel,k} = m + \left(\sigma_s+s_f{\pi}^3_s\right) \
\frac{ \left(1- |q_f B|/\Lambda^2\right)^{k+\frac{\lambda s_{\! f}-1}{2}}}
{ \left(1+ |q_f B|/\Lambda^2\right)^{k+\frac{\lambda s_{\! f}+1}{2}}}
\;\exp\!\big(-{q_{\parallel}}^{2}/\Lambda^2\big)\,,
\label{eq:cons_mass_gaus}
\end{equation} 
where the notations have the following meanings: $q_{\parallel}=(q_3,q_4)$, $s_f=\rm sign(q_f)$, $k$ is the index for the Landau level and $\lambda=\pm1$ is for the spin.

Now it is easy to obtain the two gap equations by differentiating the above equation with respect to ${\sigma_s}$ and ${\pi}^3_{s}$ and are given in appendix~\ref{app:gap_eqns}. To deal with phenomena in the presence of a hot medium we need to introduce the temperature. That is done using the well-known Matsubara formalism. It connects the Euclidean time components to the temperature. The details of the finite temperature treatment and its implications for the present model are discussed in Ref.~\cite{Ali:2020jsy}.

The quark condensate for the individual flavour can be obtained by differentiating the action with respect to the current quark mass of the corresponding flavour,
\begin{widetext}
\begin{eqnarray}
\langle\bar\psi_f\psi_f\rangle & = & - N_c \sum_{f=u,d} \frac{2|q_f B|}{2 \pi} \int \frac{d^2q_\parallel}{(2\pi)^2} \ \Bigg\{ \frac{\left(M^{s_{\! f},f}_{q_\parallel,0}+s_fa_f\right)}{q_\parallel^2 + \left({M^{s_{\! f},f}_{q_\parallel,0}}+s_fa_f\right)^2}+ \nonumber \\
& & \sum_{k=1}^\infty \frac{ \left\{2 k |q_f B| + q_\parallel^2 +
	\left(M^{+1,f}_{q_\parallel,k}+a_f\right)\left( M^{-1,f}_{q_\parallel,k}-a_f\right)\right\}\left(
	M^{+1,f}_{q_\parallel,k} + M^{-1,f}_{q_\parallel,k} \right)}{ \left\{2 k |q_f B| + q_\parallel^2 +
	\left(M^{+1,f}_{q_\parallel,k}+a_f\right)\left( M^{-1,f}_{q_\parallel,k}-a_f\right)\right\}^2 \! \! + q_\parallel^2 \left(
	M^{+1,f}_{q_\parallel,k} - M^{-1,f}_{q_\parallel,k} +2a_f\right)^2} \Bigg\}.
\label{eq:cond}
\end{eqnarray}
\end{widetext}
It is obvious that with non-zero quark masses the integral written above is divergent in the large $p$ region. It needs to be regularised and is done following the prescription of Ref.~\cite{Pagura:2016pwr},

\begin{equation}
\langle\bar\psi_f\psi_f\rangle^{\rm reg}_{B,T}=\langle\bar\psi_f\psi_f\rangle_{B,T}-\langle\bar\psi_f\psi_f\rangle^{\rm free}_{B,T}+\langle\bar\psi_f\psi_f\rangle^{\rm free,reg}_{B,T},
\label{eq:cond_reg}
\end{equation} 
where ``free" means that there is no self-interaction. The $\langle\bar\psi_f\psi_f\rangle^{\rm free,reg}_{B,T}$ is given by
\begin{widetext}
\begin{eqnarray}
\langle\bar\psi_f\psi_f\rangle^{\rm free,reg}_{B,T}&=&\frac{N_cm^3}{4\pi^2}\Bigg[\frac{\ln\Gamma(x_f)}{x_f}-\frac{\ln(2\pi)}{2x_f}+1-\left(1-\frac{1}{2x_f}\ln x_f\right)\Bigg]+\frac{N_c\left|q_fB\right|}{\pi}\sum_{k=0}^{\infty}\alpha_k\int\frac{dq}{2\pi}\frac{m}{E^f_k\left(1+\exp[E^f_k/T]\right)},
\label{eq:cond_free_reg}
\end{eqnarray}
\end{widetext}
where $x_f=m^2/(2\left|q_fB\right|)$. If there is no magnetic field then the ``$\mathrm{free, reg}$'' term will be zero.

We compare our results with the corresponding LQCD ones~\cite{Bali:2012zg} to decide on both the strength and the form of the AMM of quarks. To do so we use the renormalised condensate defined there as,
\begin{equation}
\Sigma_{B,T}^f=\frac{2m}{{\cal N}^4}\left[\langle\bar\psi_f\psi_f\rangle^{\rm reg}_{B,T}-\langle\bar\psi_f\psi_f\rangle^{\rm reg}_{0,0}\right]+1,
\label{eq:sigma_scaled_lat}
\end{equation} 
$\cal N$ is given by ${\cal N}=(m_{\pi}F_{\pi,0})^{1/2}$ that follows from the
Gell-Mann-Oakes-Renner (GOR) relation. $F_{\pi,0}$ is the pion decay constant in the chiral limit and $m_{\pi}$, the neutral pion mass. It is noteworthy that this form is particularly devised to cancel the additive and multiplicative divergences that arise in the lattice calculation.

\section{Results}
\label{sec:res}
In this section, we will first test our model predictions by comparing them with the LQCD data on condensate averages for different strengths of the magnetic field. We estimate the condensate average in our model by considering both constant and varying strengths of AMM of quarks. For the varying strengths, we use two different forms: i) $\kappa\propto\sigma_s$ and ii) $\kappa\propto\sigma_s^2$; $\sigma_s$ is the isoscalar scalar meanfield. To explore a wide range of strengths in AMM, we have taken a proportionality constant differing in order of magnitude. We take two different values of $v$, $0.1$ and $0.01$, where $v$ is the proportionality constant $(\kappa\,=\,v\sigma_s$ or $\kappa\,=\,v\sigma_s^2)$.

After obtaining the condensate average for those different conditions and comparing them with the available LQCD data for the same, we pick up those which capture the very important IMC effect and reproduce the LQCD data as closely as possible. Along with the condensate average, the phase diagram in the $T-eB$ plane also plays an important role.

With the selected strength and form of AMM through the abovementioned procedure, we move on to the next observable, the condensate difference. It is calculated for different conditions within the ambit of the model. Then it is compared with the available LQCD data for different values of $eB$.

\subsection{Condensate average}
In this subsection, we estimate the condensate average using the model for different values of the AMM of quarks along with different forms and then compare them with the LQCD data.

\subsubsection{Constant $\kappa$}
First, we discuss the results for the constant values of $\kappa$. We choose two sets of $\kappa$ that are given in the Ref.~\cite{Fayazbakhsh:2014mca}. The two sets differ by an order of magnitude. They are $\{\kappa_u=0.29016,\kappa_d=0.35986\}\,{\mathrm{GeV}^{-1}}$ which we call set-I and $\{\kappa_u=0.00995,\kappa_d=0.07975\}\,{\mathrm{GeV}^{-1}}$ which we call set-II.\\

\paragraph*{\bf Set-I:}
\begin{figure}[!htbp]
 \includegraphics[scale=0.37]{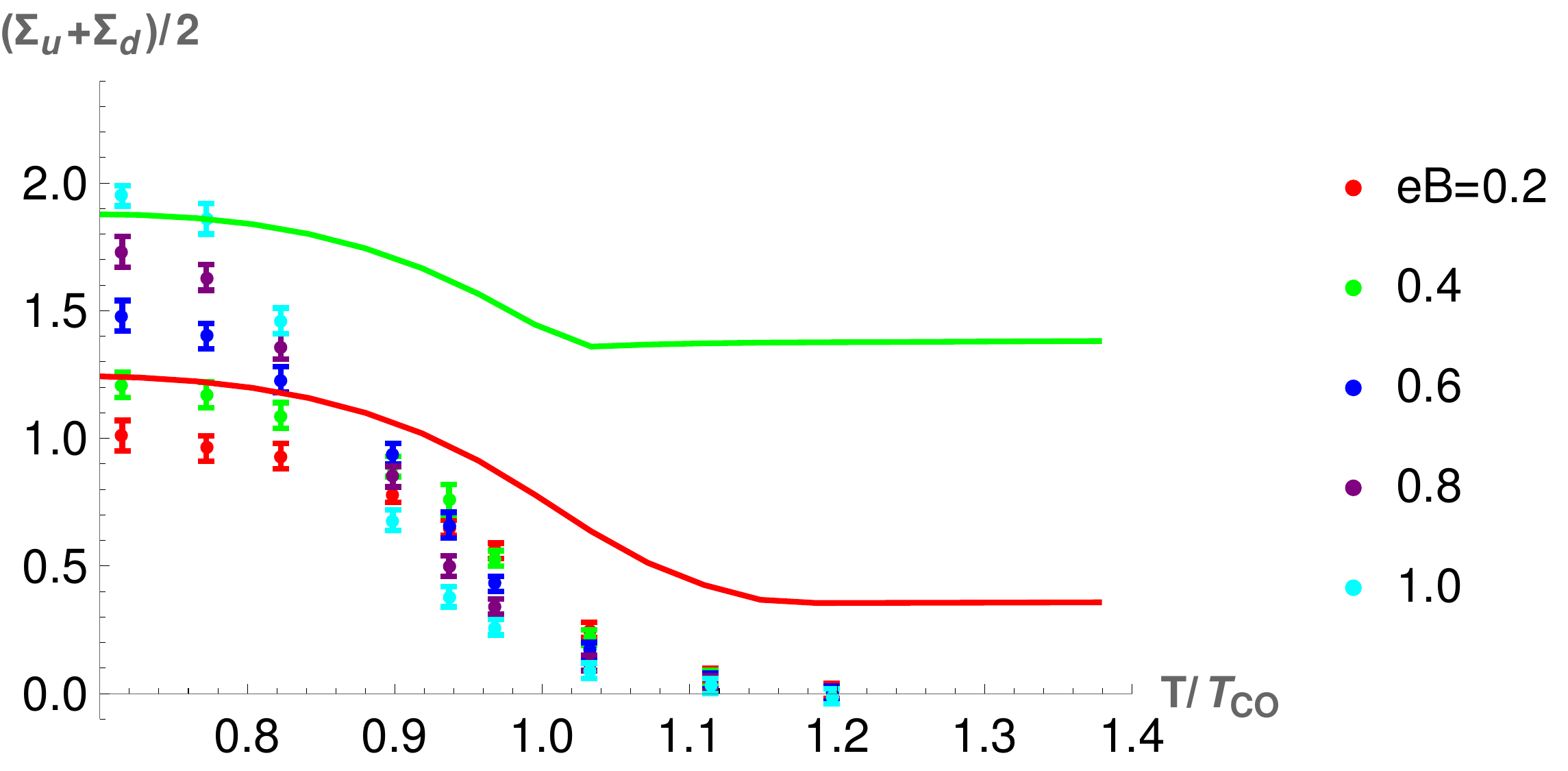}
 \caption{Averages of condensates as compared to those in LQCD data are plotted as a function of scaled temperature for five different values of the magnetic field for a constant set of $\kappa$'s (set-I). The data points are taken from the Ref.~\cite{Bali:2012zg} and the solid lines are the results from the model. Different colours symbolise different strengths of the magnetic field in a unit of $\mathrm{GeV}^2$.}
\label{fig:sigma_avg_s0_KI_c05}
\end{figure}
With the constant values of $\kappa$ from set-I the averages of the condensates are plotted as a function of temperature for different values of the magnetic field in Fig.~\ref{fig:sigma_avg_s0_KI_c05}. The corresponding LQCD data have also been shown there.

From the figure, it is clear that with these strengths of AMMs of quarks both the values and the trends of the condensate averages are not at all reliable and they are way off as compared to the lattice data as a whole. Thus, set-I is not a good choice for AMM of quarks which can be used in such effective model calculation.\\

\paragraph*{\bf Set-II:}
\begin{figure}[!htbp]
 \includegraphics[scale=0.35]{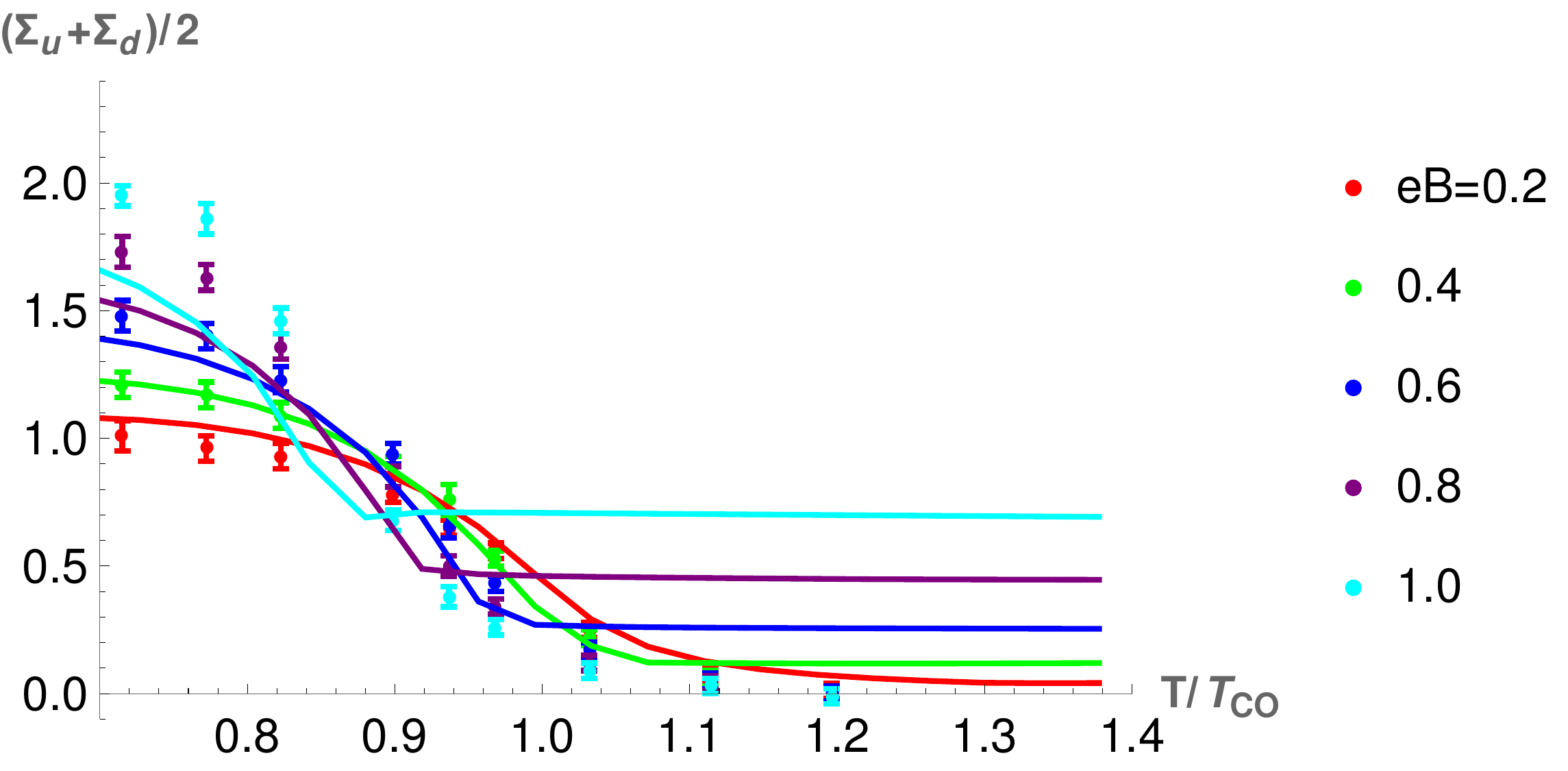}
 \caption{The same plot as in Fig.~\ref{fig:sigma_avg_s0_KI_c05} but with $\kappa$'s from set-II. Lattice data are from Ref.~\cite{Bali:2012zg}. Different colours represent different strengths of the magnetic field in a unit of $\mathrm{GeV}^2$.}
\label{fig:sigma_avg_s0_KII_c05}
\end{figure}
In Fig.~\ref{fig:sigma_avg_s0_KII_c05}, we show the plots for condensate averages for the $\kappa$ values from set-II. From the figure, we observe that the IMC effect is more or less captured for all possible strengths of the magnetic field. But the values of the condensate averages do not match with the corresponding LQCD data, particularly for the higher values of $eB$. Also, the condensate with magnetic field values higher than $0.6\, \mathrm{GeV^2}$ behaves in a peculiar manner, particularly close to the transition temperature.

%Though we can consider that the condensate average does fairly well up to a magnetic field value of $0.6\, \mathrm{GeV^2}$, but the condensate difference is well off from the lattice data and is not even reliable for higher values of $eB$. 

Thus, we can conclude here that neither set-I nor II can reliably reproduce the LQCD data for the condensate average. So, we are no more going to consider them for the rest of our analysis. The results obtained so far, indicate that the use of constant $\kappa$, maybe, is not suitable for use. In hindsight, one might find this conclusion as expected. Because we know that the AMMs of quarks are dynamically generated through the spontaneous breaking of chiral symmetry. With temperature and magnetic field, the chiral symmetry evolves and gets restored at high enough temperature. Thus, the AMMs of quarks should also evolve with temperature and magnetic field. And this is not the case with a constant $\kappa$.

In the next two sections, we explore the idea of varying the strengths of AMMs of quarks as a function of temperature. We achieve that by considering the strength of the AMM depending on the isoscalar scalar meanfield, $\sigma_s$. In the next section, we investigate the functional dependence $\kappa\propto\sigma_s$ and $\kappa\propto\sigma_s^2$, in the latter section.

\subsubsection{$\kappa\propto\sigma_s$}
In this section, we explore the idea of varying AMM of quark by assuming it as proportional to the isoscalar scalar meanfield $(\kappa\propto\sigma_s)$. Then immediately, one needs to think about the value of the proportionality constant, which we denote as $v$ $(\kappa\,=\,v\sigma_s)$. 
We now take values of $v$ differing in order of magnitude, so that we can explore a wide range of the strength of the AMM.\\

\paragraph*{$\boldsymbol{v=0.1:}$}
\begin{figure}[!htbp]
 \includegraphics[scale=0.35]{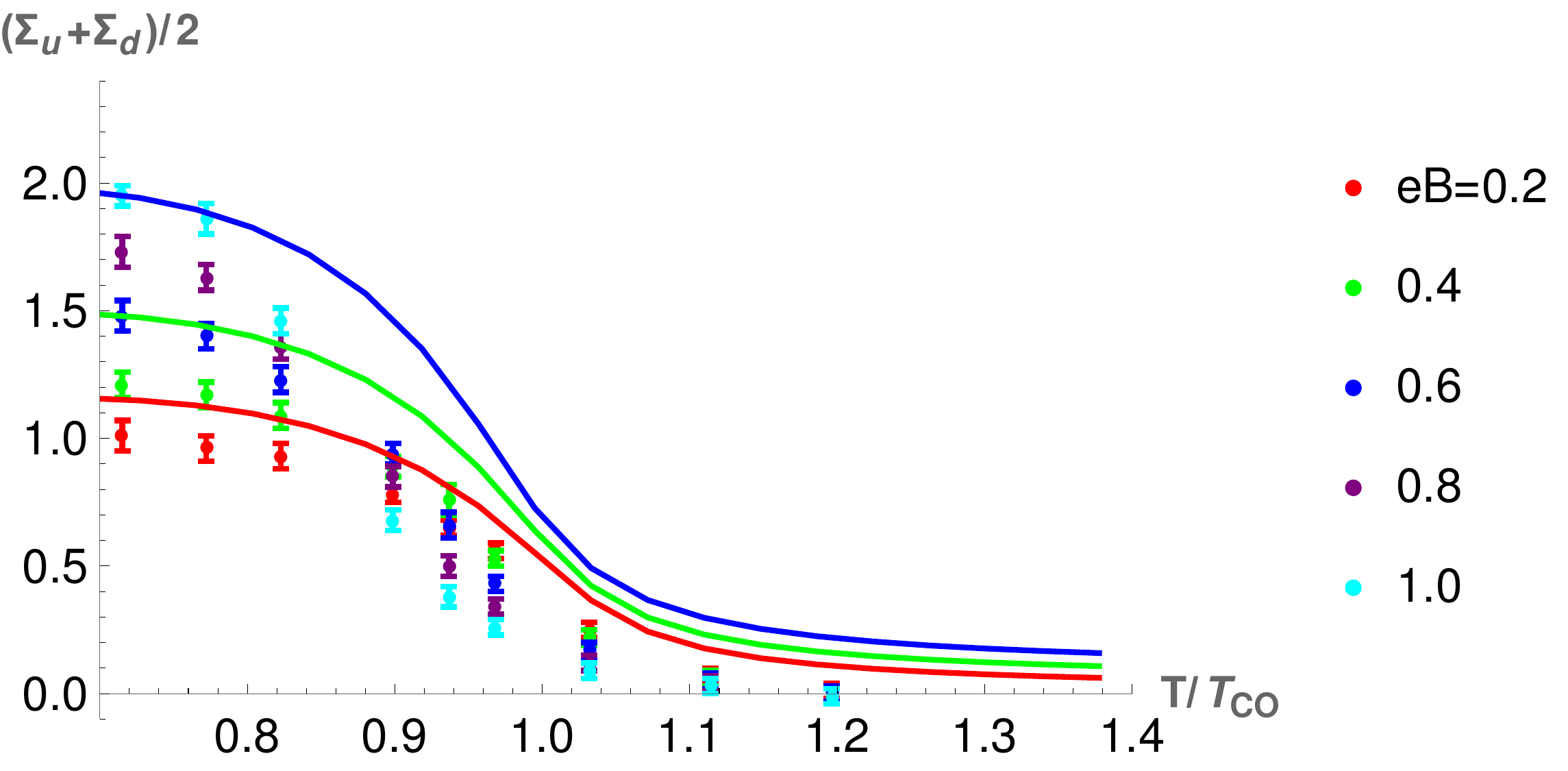}
 \caption{Averages of condensates as compared to those in LQCD data (\cite{Bali:2012zg}) are plotted for $\kappa_f$ being proportional to $\sigma_{s,f}$, with the proportionality constant $v=0.1$. The meanings of the legend are the same as in the previous figures.}
\label{fig:sigma_avg_s1_v01_c05}
\end{figure}
We begin the exploration of the $\sigma_s$ dependence of the AMM by considering $v=0.1$. We should make it clear here that we choose functional dependence for individual flavour, i.e., $\kappa_f\,=0.1\,\sigma_{s,f}$; $f$ stands for the flavour, $u$ and $d$ in our case. 

The plot for condensate averages for different values of the magnetic field is shown in Fig.~\ref{fig:sigma_avg_s1_v01_c05} along with the corresponding LQCD data. On comparing with the LQCD data, we observe that not only the strength of the condensate for a given value of a magnetic field that fails to match with the corresponding lattice data (especially for higher values of magnetic fields) but also it vehemently fails to reproduce the effect of IMC. Thus, from this observation, it can be concluded that with this form such strength of AMM is not suitable to be used in this kind of effective model. 

\paragraph*{$\boldsymbol{v=0.01:}$}
\begin{figure}[!htbp]
 \includegraphics[scale=0.35]{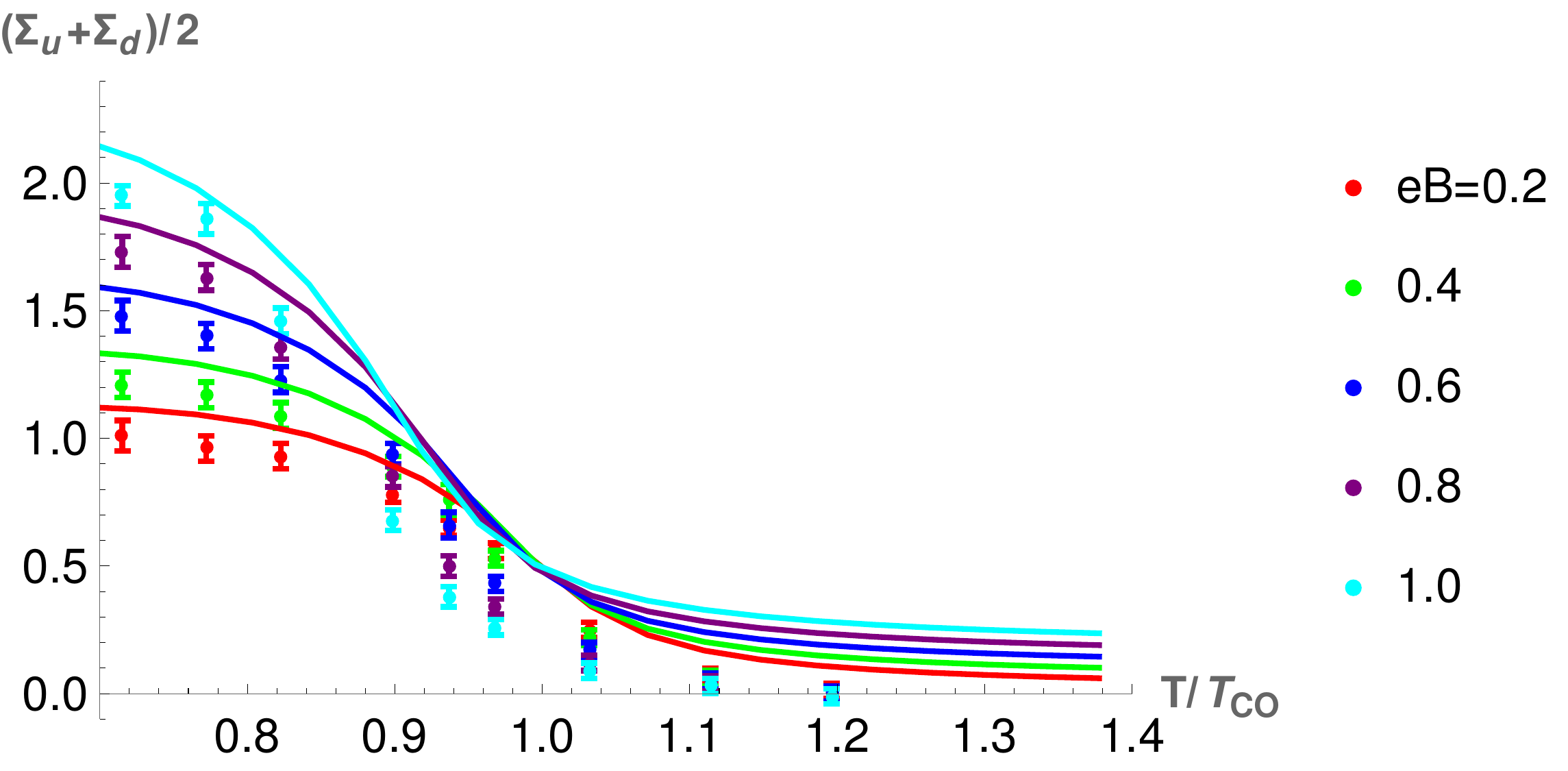}\hspace{0.5cm}
 \includegraphics[scale=0.35]{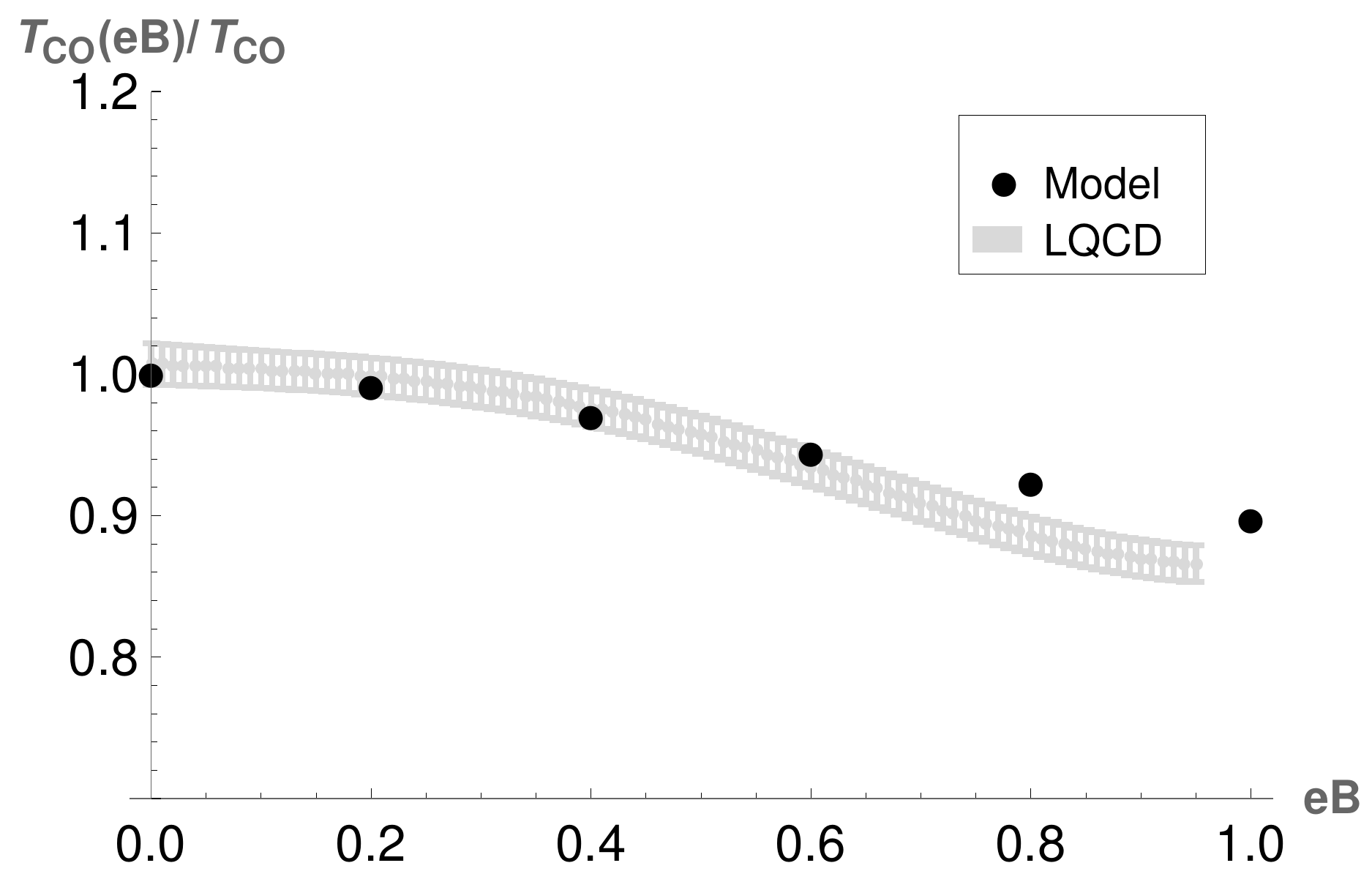}
 \caption{Left panel: condensate averages as compared to those in LQCD data (\cite{Bali:2012zg}) are plotted for $\kappa_f$ being proportional to $\sigma_{s,f}$, with the proportionality constant $v=0.01$. The meanings of the legends are the same as in Fig.~\ref{fig:sigma_avg_s0_KI_c05}. Right panel: crossover temperature calculated at different values of $eB$ within both model and LQCD~\cite{Bali:2012zg} treatment.}
\label{fig:sigma_avg_s1_v001_c05}
\end{figure}
Fig.~\ref{fig:sigma_avg_s1_v001_c05} displays the plot of condensate averages for $v=0.01$. In comparison with the available LQCD data, we find the model predictions to be quite reasonable. The strengths are within the ballpark of data for all values of the magnetic field. Also, notably, there is a hint of the presence of the IMC effect, which is another important criterion for a strength and a form of AMM to be qualified as a viable choice.

Thus, $\kappa_f$ being proportional to the corresponding meanfield, $\sigma_{s,f}$ with the proportionality constant, $v\,=\,0.01$ is definitely a possibility. It needs to be tested further by checking whether it can reproduce other quantities successfully.

\subsubsection{$\kappa\propto\sigma_s^2$}
In this subsection, we explore the possibility of AMM being proportional to the square of the meanfield $\kappa\propto\sigma_s^2$. Like the exploration in the previous subsection, here also we take two different values of the proportionality constant, $v\,=\,0.1$ and $0.01$. First, we check the result with $v=0.1$.

\paragraph*{$\boldsymbol{v=0.1:}$}
\begin{figure}[!htbp]
 \includegraphics[scale=0.35]{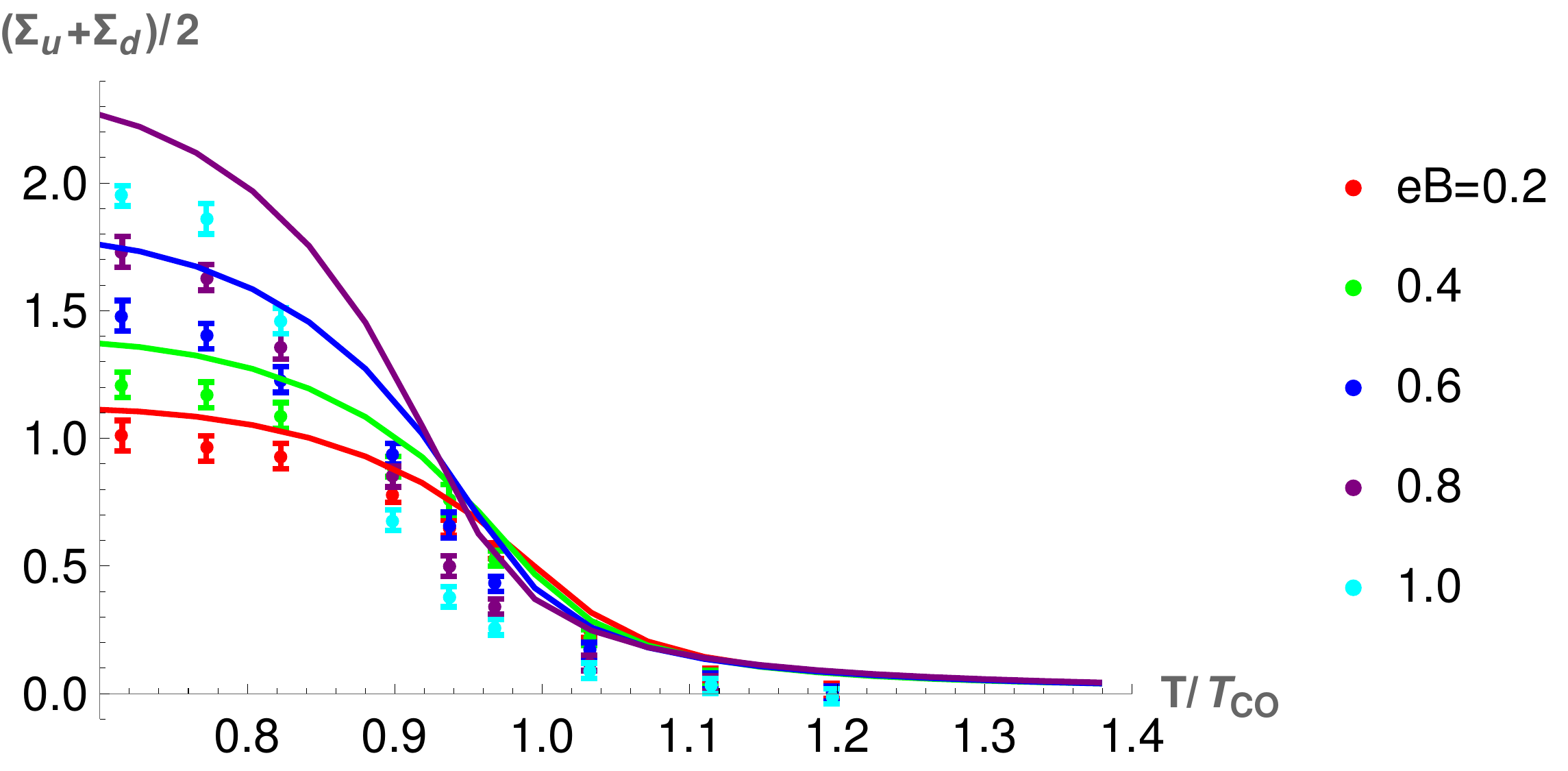}
 \caption{Averages of condensates as compared to those in LQCD data (\cite{Bali:2012zg}) are plotted for $\kappa_f$ being proportional to $\sigma_{s,f}^2$, with the proportionality constant $v=0.1$. The legends stand for the same meaning as in Fig.~\ref{fig:sigma_avg_s0_KI_c05}.}
\label{fig:sigma_avg_s2_v01_c05}
\end{figure}
In Fig.~\ref{fig:sigma_avg_s2_v01_c05}, we display the condensate average for $v=0.1$. We learn from the figure that with this set of choices for the AMM a hint of IMC effect is present there. But as we increase the magnetic field the strength of the condensate average falls apart from the corresponding lattice data. Our purpose is not only to get the IMC effect but also to match the strength of the condensates given by the LQCD data or at least to be in the ballpark. Thus, we exclude this strength of the AMM with this form as a viable choice for our model.

Next, we go to one order of magnitude lower strength of AMM with the same form.

\paragraph*{$\boldsymbol{v=0.01:}$}
\begin{figure}[!htbp]
 \includegraphics[scale=0.35]{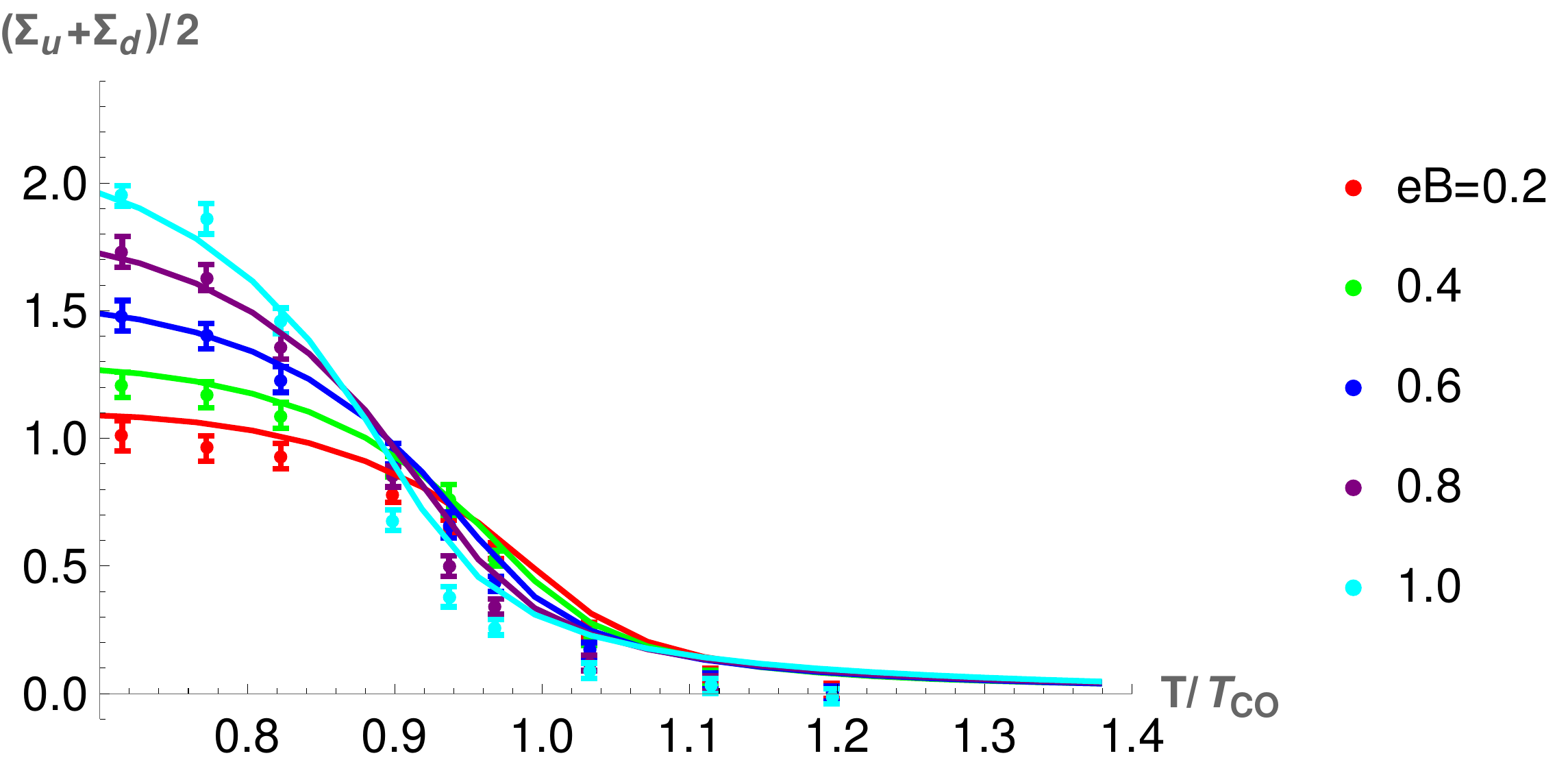}\hspace{0.5cm}
 \includegraphics[scale=0.35]{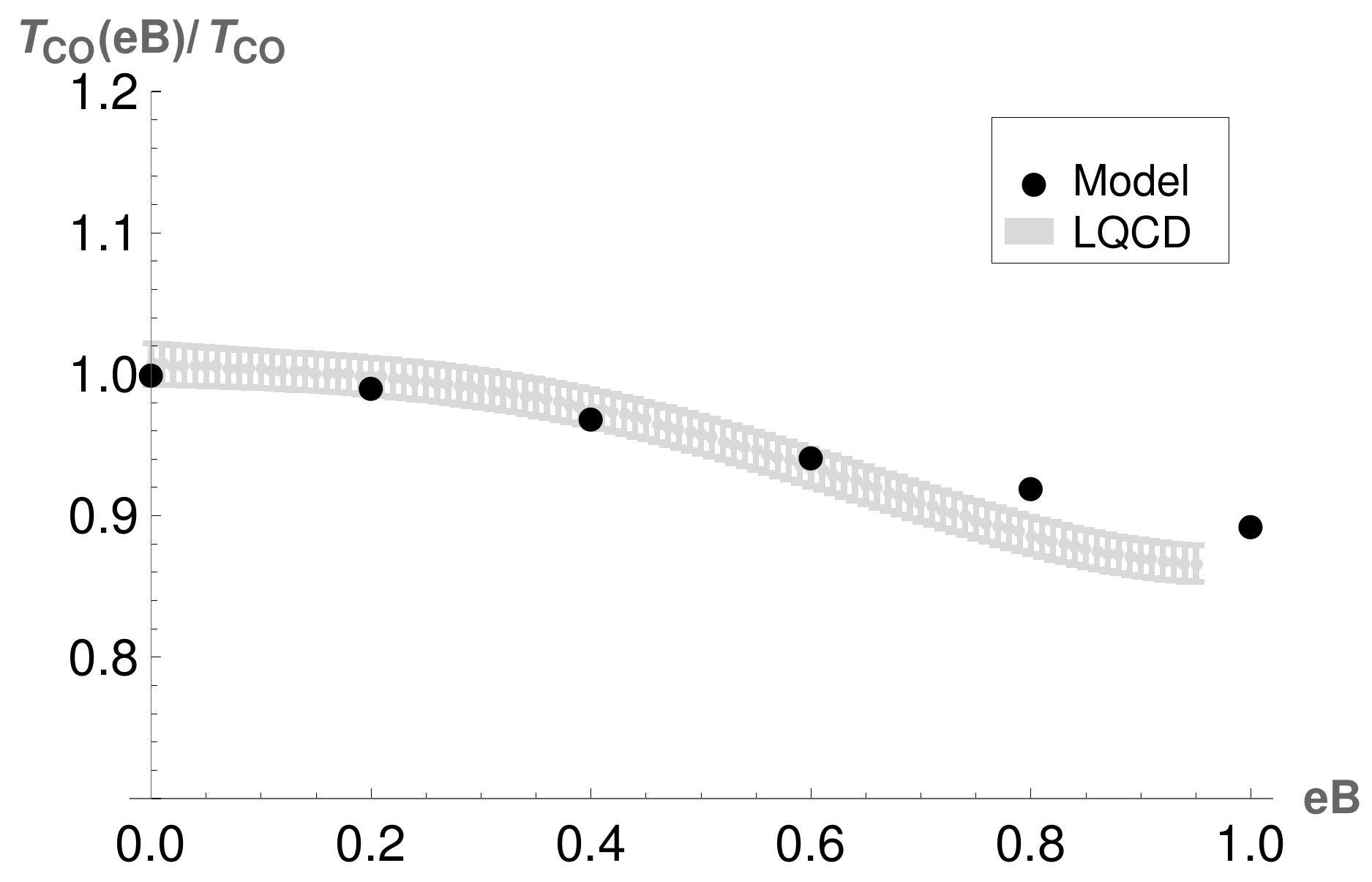}
 \caption{Left panel: averages of condensates (solid lines are results from the model) as compared to those in LQCD result (data points~\cite{Bali:2012zg}) are plotted for $\kappa_f$ being proportional to $\sigma_{s,f}^2$, with the proportionality constant $v=0.01$. Different colours stand for different strengths of $eB$. Right panel: comparison of crossover temperatures between model and LQCD estimation.}
\label{fig:sigma_avg_s2_v001_c05}
\end{figure}
In Fig.~\ref{fig:sigma_avg_s2_v001_c05}, the condensate averages are shown for $v=0.01$ with $\kappa_f$ being proportional to $\sigma_{s,f}^2$. What we observe is that first of all, the very necessary feature of IMC effect is captured with this form and strength. Secondly, the model predictions for the strength of the condensate averages are fairly comparable to that of lattice calculation. Thus, this form, $\kappa_f\propto\sigma_{s,f}^2$, with $v=0.01$ is possibly a good choice.

In the next section, we will discuss the condensate differences calculated and compared with the lattice data for different values of $eB$.  

\subsection{Condensate difference}
In the previous section, we studied the condensate averages and by comparing them with the LQCD data we decided on the set of values and forms of the AMM which could successfully reproduce the data. Out of all the sets that we explored, we found two sets to be promising. 

These are the ones which capture the IMC effect as well as are more or less comparable with the data for the whole available ranges of temperature and magnetic fields. One is the $v=0.01$ with the form $\kappa_f\propto\sigma_{s,f}$ and the other is $v=0.01$ with the form $\kappa_f\propto\sigma_{s,f}^2$.

We further test those two sets by calculating the condensate differences and comparing them with the lattice data. 
 
\subsubsection{$\kappa\propto\sigma_s\,\mathrm{with}\,v=0.01$}
\begin{figure}[!htbp]
 \includegraphics[scale=0.4]{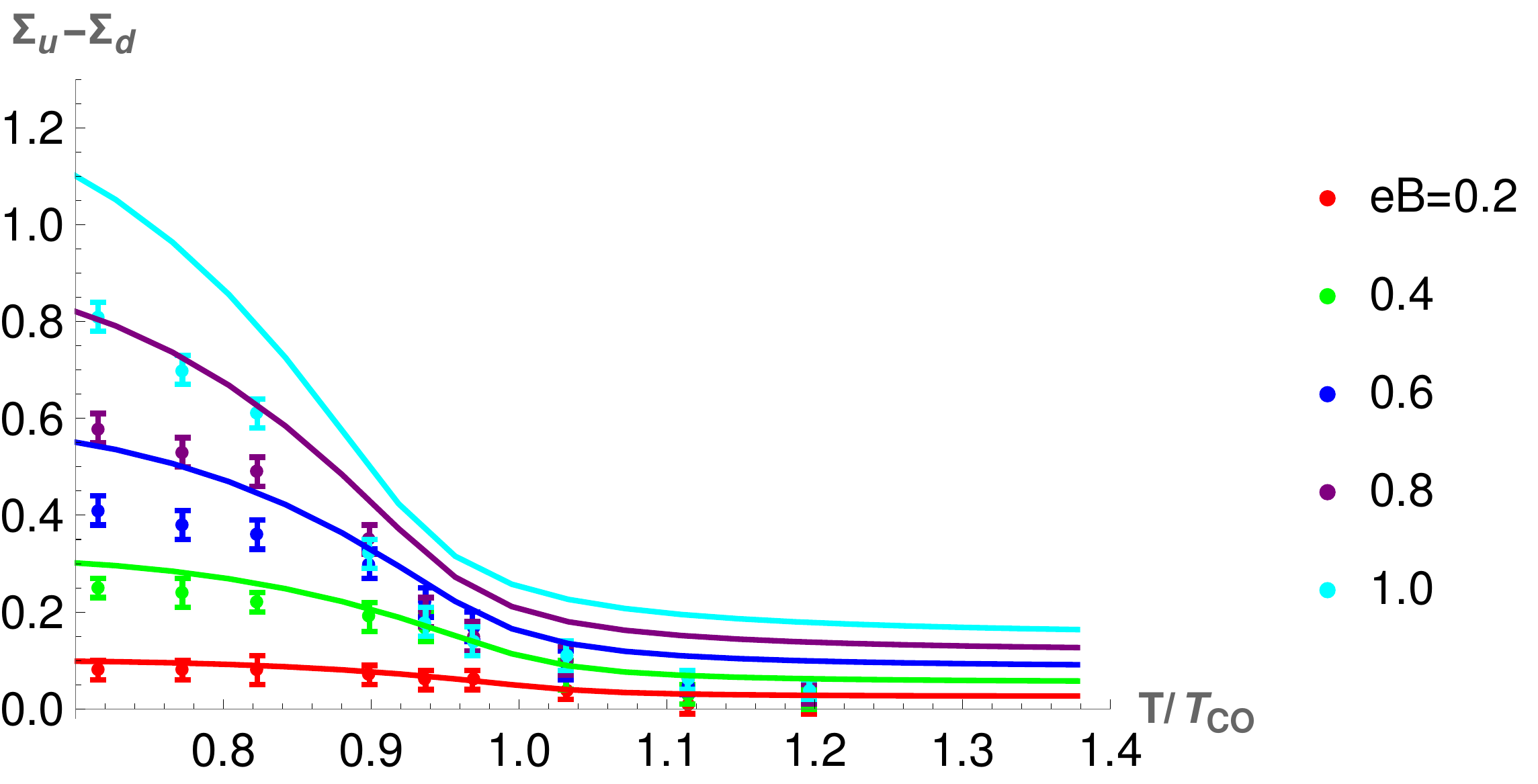}
 \caption{Differences between up and down condensates as compared to those in LQCD data (\cite{Bali:2012zg}) are plotted for $\kappa_f$ being proportional to $\sigma_{s,f}$, with the proportionality constant $v=0.01$. The points represent the lattice data and the solid lines are model predictions. The magnetic field is in the unit of $\mathrm{GeV}^2$.}
\label{fig:sigma_dif_com_s1_v001_c05}
\end{figure}
In Fig.~\ref{fig:sigma_dif_com_s1_v001_c05}, the condensate differences are plotted for $v=0.01$ with $\kappa_f\,=\,v\sigma_{s,f}$. As we observe from the figure that the condensate differences calculated within the model are comparable with the data for lower values of the magnetic field. As we go beyond the value of $eB\,=\,0.6\, \mathrm{GeV}^2$ the model predictions no more remain comparable to lattice data. 

Thus, this form $(\kappa_f\propto\sigma_f)$ with the strength $v=0.01$, though could reliably reproduce the condensate average, cannot successfully reproduce the differences for all values of the magnetic field. Next, we investigate the other form of the AMM with $v=0.01$.

\subsubsection{$\kappa\propto\sigma_s^2\,\mathrm{with}\,v=0.01$}
\begin{figure}[!htbp]
 \includegraphics[scale=0.4]{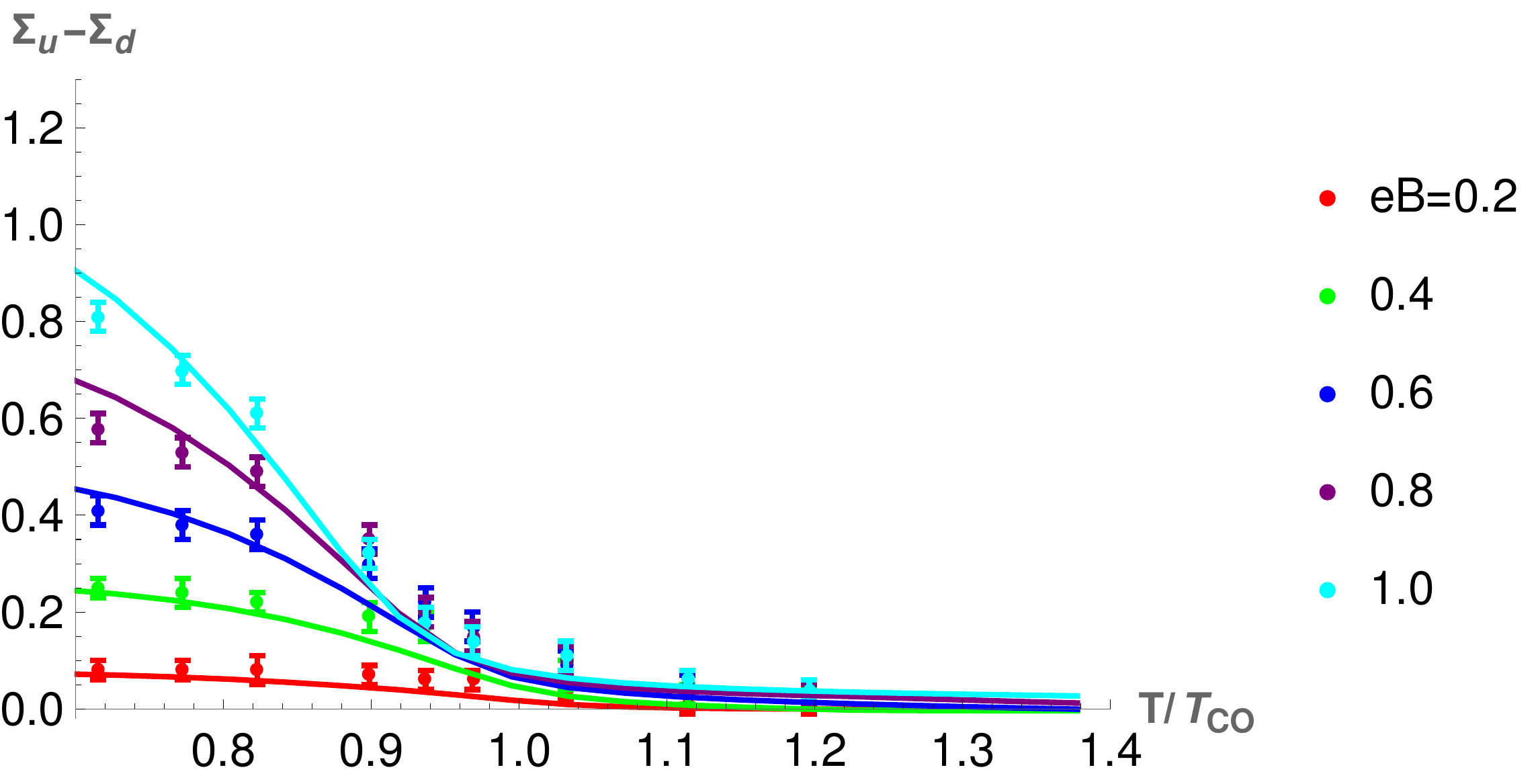}
 \caption{Differences of condensates as compared to those in LQCD result are plotted for $\kappa_f$ being proportional to $\sigma_{s,f}^2$, with the proportionality constant $v=0.01$. Meanings of the legends are the same as in Fig.~\ref{fig:sigma_dif_com_s1_v001_c05}.}
\label{fig:sigma_dif_com_s2_v001_c05}
\end{figure}
We show the plots for the condensate differences in Fig.~\ref{fig:sigma_dif_com_s2_v001_c05} for all possible values of the magnetic field. On comparing with the lattice data we see that the model predictions are fairly reasonable for all values of $eB$ throughout the temperature range. Thus, we found out that $\kappa_f\propto\sigma_{s,f}^2$ with $v=0.01$ could successfully reproduce not only the condensate averages but also the condensate differences as predicted by the LQCD study. It can as well reproduce the lattice predicted phase diagram in the $T-eB$ plane.

\section{Conclusion}
\label{sec:con}
Perturbative calculation predicts that massless quarks cannot have a measurable anomalous magnetic moment (AMM). On the other hand, spontaneous breaking of the chiral symmetry renders the quarks to be massive. The massive quarks can possess a measurable strength of AMM. Thus, the AMM of quarks is dynamically generated. This moment can couple to a magnetic field.

There are different possible scenarios starting from heavy-ion collisions (HIC) to magnetars, where a presence of a very strong magnetic field is predicted. Once it is produced it will interact with the strongly interacting QCD medium present in those kinds of scenarios. Thus, the understanding of AMM of quarks in presence of a magnetic field becomes crucial.

Unlike the AMM of fermions sans strong interaction (for example, leptons), the quark's AMM involves nonperturbative regimes. This ``nonperturbativeness'' renders the study extremely difficult using known analytical techniques. Here comes the role of effective QCD models, which are often exploited to understand different nonperturbative aspects of QCD. One such model is the Nambu\textemdash Jona-Lasinio (NJL) model, which is very successful in describing different properties of QCD.

It has also been used on numerous occasions to explore the AMM of quarks in presence of a strong magnetic field~\cite{Fayazbakhsh:2014mca,Chaudhuri:2019lbw,Chaudhuri:2020lga,Xu:2020yag,Mei:2020jzn,Aguirre:2020tiy,Chaudhuri:2021skc,Ghosh:2021dlo,Aguirre:2021ljk,Wen:2021mgm,Farias:2021fci,Wang:2021dcy,Chao:2022bbv,Mao:2022dqn,He:2022inw,Qiu:2023kwv}. In most of the articles, the presence of AMM has been exploited to understand the effect of magnetic catalysis (MC) and inverse magnetic catalysis (IMC). Along with the phase diagram of QCD in the $T-eB$ plane has also been explored. Quantities like the dilepton rate have also been explored in presence of AMM of quarks. 

In all these works, one thing has not been properly addressed, which is the strength of the AMM of quarks at zero temperature. Also, how it varies as a function of the temperature and the magnetic field is also not clearly known. There are some efforts in this direction~\cite{Xu:2020yag,Ghosh:2021dlo,He:2022inw,Kawaguchi:2022dbq}, but nothing very conclusive. 

In this article, we tried to answer this question by incorporating the AMM in a framework of nonlocal NJL model. The primary of advantage of using the nonlocal version over the local one is that it automatically captures the IMC effect \textemdash\, one does not need to introduce any extra parameter~\cite{Pagura:2016pwr,Ali:2020jsy}. Thus, the incorporation of AMM in this framework is not to obtain the IMC effect but rather to test the allowed range of AMM of quarks along with its form through the comparison with available lattice QCD data.

We started with both constant and varying values of AMM of quarks. The constant values are picked from the Ref.~\cite{Fayazbakhsh:2014mca}. To use the temperature dependent values we chose two different ansatzes: the AMM is proportional to the i) meanfield and ii) square of the meanfield. Though such ansatzes have been previously used in a local framework~\cite{Xu:2020yag,He:2022inw,Kawaguchi:2022dbq}. When we utilise the ansatzes in our nonlocal setup we found out that the outcome is much more conclusive. As for the proportionality constant $(v)$, we took two different values differing from each other in one order of magnitude.

We divided our analysis into two parts. First, we calculate the condensate average using our model and compare it with the same in lattice QCD calculation. This comparison helps us to eliminate many of our initial choices and we are left with fewer options as we move to the second part of our analysis. In the process, we also kept track of the $T-eB$ phase diagram as obtained in our model. We observed that constant values of AMM are not suitable at all to be used in such effective models. Both ansatzes with proportionality constant $v=0.01$ are reliable as far as the condensate averages are concerned.

In the second part of the analysis, the data for the condensate difference from the lattice QCD have been compared. This is done with the available choices left after the first part of the analysis. We noticed that the ansatz with the AMM being proportional to the square of the meanfield is the most reliable, which is also in agreement with the result obtained in Ref.\cite{Kawaguchi:2022dbq}.

We remind our readers that the most important goal of this study was to investigate a reliable functional form (ansatz) for AMM of quarks. We believe that we have been able to do that using a nonlocal framework. Though the preferable value of $v$ remains as $0.01$, one should understand that this value is not very strict in nature and can be varied within a certain range to obtain an acceptable outcome. This is particularly because the framework itself does not give much scope for such a quantitative conclusion. Finally, we think that with this encouraging outcome, we can further possibly calculate some more quantities such as magnetic susceptibility, which will be reported elsewhere.

\section{Acknowledgements}
\label{sec:acknow}
C.A.I. would like to acknowledge the financial support by the Chinese Academy of Sciences President's International Fellowship Initiative under Grant No. 2020PM0064. M.S.A. would like to acknowledge the support provided by TIFR, Mumbai where the initial part of the work was done, and would also like to acknowledge the support from NISER for the current position of a Senior Project Associate. M.H. is supported by the National Natural Science Foundation of China (NSFC) Grant Nos:12235016, 12221005, 11725523, 11735007, 12275108, and the Strategic Priority Research Program of Chinese Academy of Sciences under Grant Nos XDB34030000, the start-up funding from University of Chinese Academy of Sciences(UCAS), and the Fundamental Research Funds for the Central Universities.

\appendix
\section{Gap equations}
\label{app:gap_eqns}

The two gap equations can be obtained by differentiating Eq.~\ref{eq:s_mfa} with respect to ${\sigma_s}$ and ${\pi}^3_{s}$ as,
\begin{widetext}
\begin{eqnarray}
\frac{\partial\Omega}{\partial\sigma_s} & = & \frac{ 
	\sigma_s}{G_0} - N_c \sum_{f=u,d} \frac{2|q_f B|}{2 \pi} \int \frac{d^2q_\parallel}{(2\pi)^2} \Bigg\{ \frac{\left({M^{s_{\! f},f}_{q_\parallel,0}}+s_fa_f\right){A^{s_{\! f},f}_{q_\parallel,0}\,}}{q_\parallel^2 + \left({M^{s_{\! f},f}_{q_\parallel,0}}+s_fa_f\right)^2}+ \nonumber \\
&& \sum_{k=1}^\infty\Bigg[ \frac{ \left( 2 k |q_f B| + q_\parallel^2 +
	\left(M^{+1,f}_{q_\parallel,k}+a_f\right)\left( M^{-1,f}_{q_\parallel,k}-a_f\right)\right)\left(
	A^{+1,f}_{q_\parallel,k}\left( M^{-1,f}_{q_\parallel,k}-a_f\right) + \left(M^{+1,f}_{q_\parallel,k}+a_f\right)A^{-1,f}_{q_\parallel,k} \right)}{ \left( 2 k |q_f B| + q_\parallel^2 +
	\left(M^{+1,f}_{q_\parallel,k}+a_f\right)\left( M^{-1,f}_{q_\parallel,k}-a_f\right)\right)^2  + q_\parallel^2 \left(
	M^{+1,f}_{q_\parallel,k} - M^{-1,f}_{q_\parallel,k} +2a_f\right)^2}\nonumber \\
&& +\frac{  q_\parallel^2 \left(
	M^{+1,f}_{q_\parallel,k} - M^{-1,f}_{q_\parallel,k}+2a_f \right)\left(
	A^{+1,f}_{q_\parallel,k} - A^{-1,f}_{q_\parallel,k} \right)}{ \left( 2 k |q_f B| + q_\parallel^2 +
	\left(M^{+1,f}_{q_\parallel,k}+a_f\right)\left( M^{-1,f}_{q_\parallel,k}-a_f\right)\right)^2  + q_\parallel^2 \left(
	M^{+1,f}_{q_\parallel,k} - M^{-1,f}_{q_\parallel,k} +2a_f\right)^2}\Bigg] \Bigg\}=0\,\, \mathrm{and}
\label{eq:gap_one}
\end{eqnarray}
\begin{eqnarray}
\frac{\partial\Omega}{\partial\pi^3_s} & = & \frac{\pi^3_s}{(1-2c)G_0} - N_c \sum_{f=u,d} s_f\frac{2|q_f B|}{2 \pi} \int \frac{d^2q_\parallel}{(2\pi)^2} \Bigg\{ \frac{\left({M^{s_{\! f},f}_{q_\parallel,0}}+s_fa_f\right){A^{s_{\! f},f}_{q_\parallel,0}\,}}{q_\parallel^2 + \left({M^{s_{\! f},f}_{q_\parallel,0}}+s_fa_f\right)^2}+ \nonumber \\
&& \sum_{k=1}^\infty\Bigg[ \frac{ \left( 2 k |q_f B| + q_\parallel^2 +
	\left(M^{+1,f}_{q_\parallel,k}+a_f\right)\left( M^{-1,f}_{q_\parallel,k}-a_f\right)\right)\left(
	A^{+1,f}_{q_\parallel,k}\left( M^{-1,f}_{q_\parallel,k}-a_f\right) + \left(M^{+1,f}_{q_\parallel,k}+a_f\right)A^{-1,f}_{q_\parallel,k} \right)}{ \left( 2 k |q_f B| + q_\parallel^2 +
	\left(M^{+1,f}_{q_\parallel,k}+a_f\right)\left( M^{-1,f}_{q_\parallel,k}-a_f\right)\right)^2  + q_\parallel^2 \left(
	M^{+1,f}_{q_\parallel,k} - M^{-1,f}_{q_\parallel,k} +2a_f\right)^2}\nonumber \\
&& +\frac{  q_\parallel^2 \left(
	M^{+1,f}_{q_\parallel,k} - M^{-1,f}_{q_\parallel,k}+2a_f \right)\left(
	A^{+1,f}_{q_\parallel,k} - A^{-1,f}_{q_\parallel,k} \right)}{ \left( 2 k |q_f B| + q_\parallel^2 +
	\left(M^{+1,f}_{q_\parallel,k}+a_f\right)\left( M^{-1,f}_{q_\parallel,k}-a_f\right)\right)^2  + q_\parallel^2 \left(
	M^{+1,f}_{q_\parallel,k} - M^{-1,f}_{q_\parallel,k} +2a_f\right)^2}\Bigg] \Bigg\}=0,
\label{eq:gap_two}
\end{eqnarray}
\end{widetext}
respectively. The quantity $A^{\lambda,f}_{q_\parallel,k} $ is given by

\begin{equation}
A^{\lambda,f}_{q_\parallel,k} = \frac{ \left(1- |q_f B|/\Lambda^2\right)^{k+\frac{\lambda s_{\! f}-1}{2}}}
{ \left(1+ |q_f B|/\Lambda^2\right)^{k+\frac{\lambda s_{\! f}+1}{2}}}
\;\exp\!\big(-{q_{\parallel}}^{2}/\Lambda^2\big).
\label{eq:nonlocality_form}
\end{equation} 

\bibliography{ref}

\end{document}